\documentstyle[prl,aps,twocolumn,epsfig]{revtex}
\begin{document}
\twocolumn[\hsize\textwidth\columnwidth\hsize\csname @twocolumnfalse\endcsname
\title{Self-averaging of Random and Thermally disordered diluted Ising systems}

\author{Manuel I. Marqu\'es and Julio A. Gonzalo}

\address {Depto. de F\'isica de Materiales C-IV
Universidad At\'onoma de Madrid
Cantoblanco 28049 Madrid Spain}

\date{\today}

\maketitle

\begin{abstract}
Self-averaging of singular thermodinamic quatities at criticality for randomly and thermaly 
diluted three dimensional Ising systems has been studied by the Monte Carlo approach.
Substantially improved self-averaging is obtained for critically clustered (critically 
thermally diluted) vacancy distributions in comparison with the observed self-averaging 
for purely random diluted distributions.Critically thermal dilution, leading to maximum 
relative self-averaging, corresponds to the case when the characteristic vacancy 
ordering temperature ($\theta$) is made equal to the magnetic critical temperature for 
the pure 3D Ising system ($T_{c}^{3D}$).For the case of a high ordering temperature
 ($\theta >> T_{c}^{3D}$), the self-averaging obtained is comparable to that in a 
randomly diluted system.\hspace{5em}\vskip1pcPACS numbers: 05.50.+q, 75.10.Nr, 75.40.Mg, 75.50.Lk
\end{abstract}
\vskip1pc]

\narrowtext


Systems with quenched randomness have been studied intensively for several
decades \cite{Domb}. One of the first results was the establishment of
Harris criterion \cite{Harris}, which predicts that a weak dilution does not
change the character of the critical behavior near second order phase
transitions for sustems of dimension $d$ with specific heat exponent lower
than zero (so called P systems), $\alpha _{pure}<0\Longrightarrow \nu
_{pure}>2/d$, in the nonrandom case. This criterion has been supported by
several renormalization group (RG) analyses \cite
{Lubensky,LubenskyII,Grinstein}, and by scaling analysis \cite{Aharony}. It
was shown to hold also with strong dilution by Chayes et al. \cite{Chayes}.
For $\alpha _{pure}>0$ (called R systems), for example the Ising 3D case,
the system fixed point flows from a pure (undiluted) fixed point towards a
new stable fixed point at which $\alpha _{pure}<0$ \cite
{Lubensky,LubenskyII,Grinstein,Aharony,Chayes,Kinzel}.

For a random hypercubic sample of linear dimension $L$ and a number of sites 
$N=L^d$, any observable singular property $X$ presents different values for
the different realizations of randomness corresponding to the same dilution.
This means that X behaves as a atochastic variable with average $[X]$,
variance $(\Delta X)^2$ and a normalized square width $R_X=(\Delta
X)^2/[X]^2 $. A system is said to exhibit self averaging (SA) if $%
R_X\rightarrow 0$ as $L\rightarrow \infty $. If the system is away from
criticality, $L>>\xi $, being $\xi $ the correlation lenght, the central
limit theorem indicates that strong SA must be expected. Hovewer, the
behavior of a ferromagnet at criticality, where $\xi >>L$, is not so
obvious. This point has been studied recently. Wiseman and Domany (WD)
investigated the self-averaging of diluted ferromagnets at criticality by
means of finite-size scaling calculations\cite{Wiseman}, concluding weak SA
for both the P and R cases. In contrast Aharony and Harris (AH), using a
renormalization group analysis in $d=4-\varepsilon $ dimensions, proved the
expectation of a rigorous absence of self-averaging in critically random
ferromagnets \cite{AharonyII}. More recently, WD used Monte Carlo
simulations to check the lack of self-averaging in critically disordered
magnetic systems \cite{WisemanII}. The absence of self-averaging was
confirmed. The source of discrepancy with their previous scaling analysis
was attributed to the particular size dependence of the distribution of
pseudocritical temperatures used in their work.

Quenched randomness has been investigated basically under two different
constrains, a grand-canonical constrain (average density of occupied spin
states $(p)$ fixed) and a canonical constrain (total number of occupied
sites $(c)$, constant). The appicability of the result obtined by Chayes et
al. also to canonical ensembles, implying universality between both kinds of
constrains, has been recently investigated \cite{Pazmandi,AharonyIII}. Monte
Carlo simulations \cite{WisemanII,AharonyIII} indicate universality for $R_X$
in the grand-canonical constrain ensembles with different concentrations $%
(p) $. Universality between results obtained in the canonical and the
grand-canonical constrains at very large values of L is implied by the work
of Aharony et al.\cite{AharonyIII}.

In all cases previously investigated frozen disorder was always produced in
a random way, that is, vacancies were distributed throughout the lattice
randomly. Real systems, however, can be realized in other kinds of vacancy
distributions. In particular, dilution in a ferromagnetic lattice can be
produced by an equilibrium thermal order- disorder distribution of
vacancies, governed by a characteristic ordering temperature ($\theta $). If
this ordering temperature $\theta $ is high enough, the equilibrium thermal
disorder will be similar to the random disorder of previous investigations.
However, if $\theta $ happens to coincide with the characteristic magnetic
critical temperature ($T_c=4.511617$) of the undiluted system, new
possibilities are open. In the present work we study the effects on the
magnetization self-averaging in diluted ferromagnets obtained by thermal
vacancy distributions in three dimensional Ising systems using the Monte
Carlo approach. We will study the self-averaging or lack of it in these
systems and will compare the results with those obtained with random vacancy
distributions for the same concentration. In order to actually produce these
so called thermally diluted systems we make the following steps: First we
thermalize the pure system at a given temperature $\theta $, finding a
number N+ of (+) sites and a number N- of (-) sites in thermal equilibrium.
Then we consider the minority kind, either (+) or (-) and we label them as
vacancies, thereafter frozening in the disordered system. Only sites of the
majority kind will be occupied by spins. Once the spin system has been
prepared in this way, we can study any singular magnetic physical property
at a given temperature ($T$).

The spin system (i) so constructed will have a magnetic sites concentration $%
c_i=\left| N_s\right| /N$, where $N_s$ is the number of majority spins from
the pure system and $N=N_{+}+N_{-}$ is the total number of sites. For a
large enough value of $L$ and $\theta \geq T_c^{3D}$, $c_i$ will approach
0.5 in most cases.

We perform Monte Carlo calculations of the magnetization per spin at
criticality for systems thermally and randomly diluted, using in both cases
the Wolff \cite{Wolff} single cluster algorithm \cite{Wang} with periodic
boundary conditions, on lattices of different sizes $L$. Determinations of
the magnetic critical temperature for different dilutions ($T_c(p)$) were in
good agreement with those previously given with high accuracy by Heuer \cite
{Heuer} and by Wiseman et al. \cite{WisemanII}.

We consider thermally diluted samples at two different values of $\theta $, (%
$\theta =4.5115=T_c^{3D}$, and $\theta =1000>>T_c^{3D}$) for the same value
of $L=40$. We can compare this thermal distributions of vacancies with the
equivalent random distribution with probability $p=0.5$. We will consider
the magnetization per spin for each sample at criticality, that is the
magnetization $M_i[T_c(p)]$ for each realization ''i'' within the full set
of samples. In order to show at a glance the resulting dispersion in $c$ and
in $M$, we use an scattered plot $c$ vs. $M$, where each point on the plane
represents an actual realization of a given density of spins $c_i$ and a
given magnetization per spin $M_i$. This plots contains more information
than the usual histograms.

First we consider what we may call the {\bf hipercritical} case where $%
\theta =1000$. Results are shown in Fig. 1 for $L=40$. As expected, both
diluted sets realized by thermal and random dilution look almost the same,
showing the equivalent effects of a high ordering temperature and random
dilution. Histograms for the magnetization are shown, at the
''grand-canonical'' case with $p=0.5$ and $c>0.5$ (in our realizations,
there are never more vacancies than spins). Similar normalized square widths
are found, $R_M^{random}\approx R_M^{thermal}\approx 0.06$ in both cases.
They are smaller than the ones on Ref. 13 due to the fact that for us $c>0.5$
always.

Next we consider the {\bf critical} case, $\theta =T_c^{3D}.$ Results are
shown in Fig. 2 again for $L=40$. The situation changes completely. Now the
critically thermally ordered set behaves different, with an increase on the
dispersion in $c$. The magnetization has been calculated

\begin{figure}[h]
\epsfig{width=0.95 \linewidth,figure=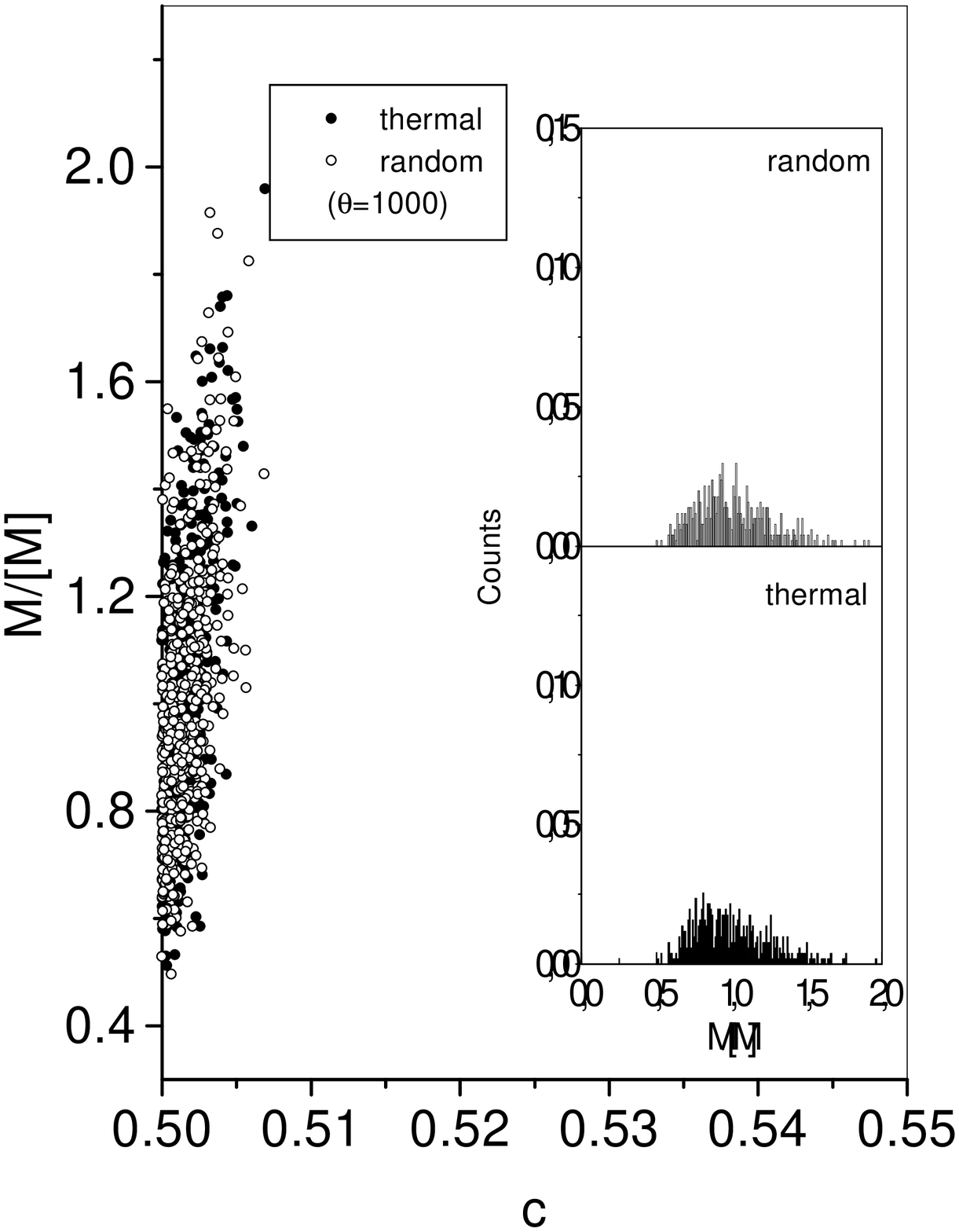} 
\vspace{0.0truecm}
\caption{Scattered plot of the concentration of spins ($c$) versus the
normalized magnetization per spin $M/[M]$ ($[M]$ is the media of the
magnetization in all the realizations: $[M]^{random}\approx
[M]^{thermal}\approx 0.15$). Results are shown for the random case with $%
p=0.5$ and $c\geq 0.5$ (open circles) and the hipercritical thermal case
with $\theta =1000>>T_c^{3D}$ (black circles). Dispersion of the
magnetization and dispersion of the concentration are similar in both cases.
Inset shows the histograms for the values of the normalized magnetization in
both cases. The upper case corresponds to random dilution and the lower to
thermal dilution. Note the equivalence of the shapes resulting in similar
values for the normalized square widths.}
\label{fig1}
\end{figure}

at the critical temperature for a dilution equal to $p=0.5$, $T_c(p=0.5)$ in
order to compare the normalized square width in both cases. In this case we
find extremely low values of the normalized square width for the thermally
diluted set of samples in the ''grand-canonical'' case, $R_M^{thermal}%
\approx 10^{-3}$, which is around $60$ times smaller than the width found
for the random case. (See the width differences in the histograms on Fig. 2).

To study the self-averaging evolution we must consider the behavior of the
normalized square width $R_M$ with the length $L$ of the system.
Consequently we have diluted the systems thermally at criticality ($\theta
=T_c^{3D}$) for different values of $L$. Of course, in all this realizations
for critical thermal cases, the corresponding value of p depends on the
value of $L$, but tends always to $0.5$ as $L$ increases as it should. We
have then calculated the magnetization for $T=T_c(p=0.5)$ in order to
compare the evolution of the normalized square widths of the thermally
diluted cases under the ''grand-canonical'' constrain with the respective
widths corresponding to random dilution with $p=0.5$.

\begin{figure}[h]
\epsfig{width=0.95 \linewidth,figure=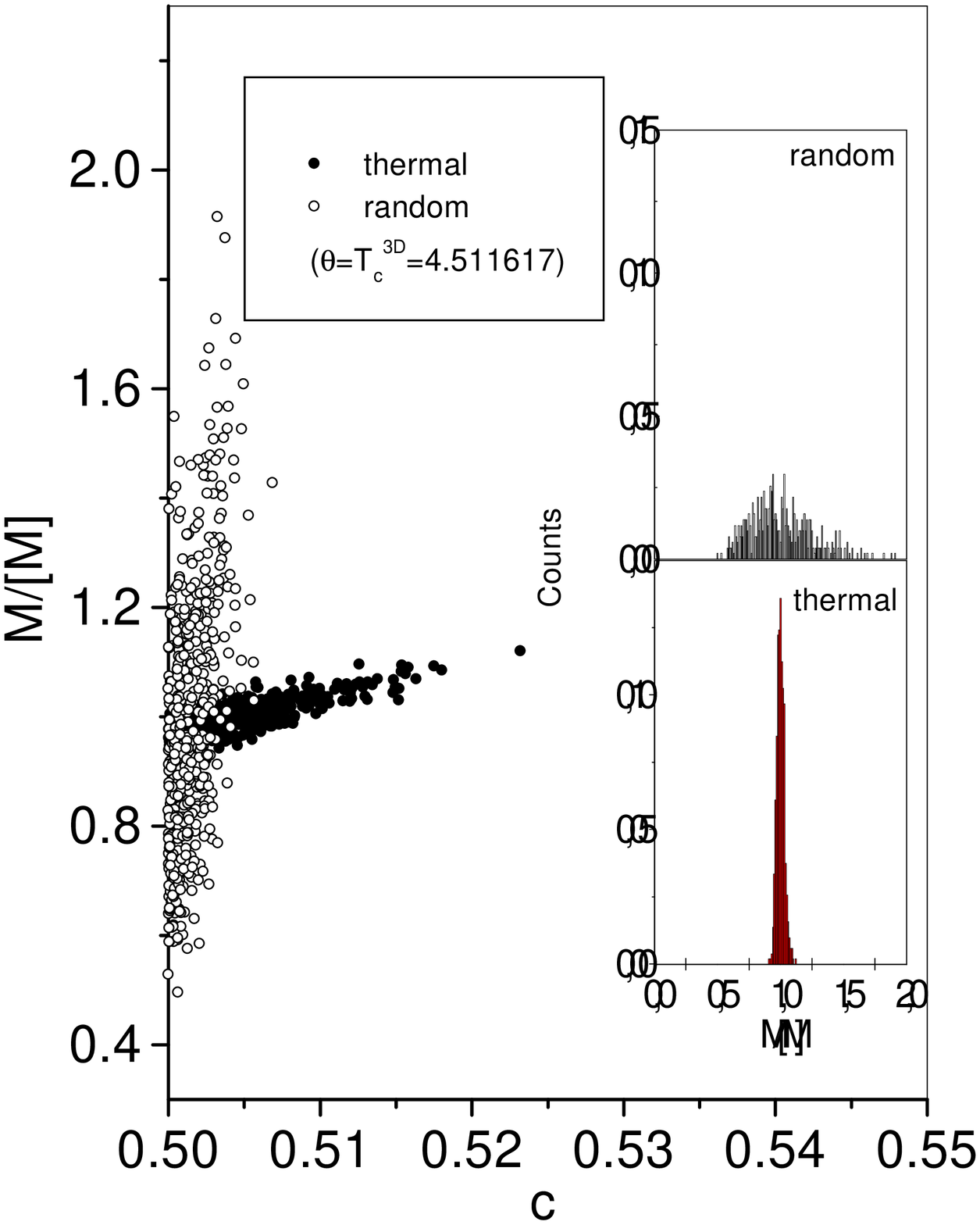} 
\vspace{0.0truecm}
\caption{Scattered plot of the concentration of spins ($c$) versus the
normalized magnetization per spin $M/[M]$ ($[M]$ is the media of the
magnetization in all the realizations: $[M]^{random}\approx 0.15,
[M]^{thermal}\approx 0.6$). Results are shown for the random case with $%
p=0.5 $ and $c\geq 0.5$ (open circles) and the critical thermal case with $%
\theta =4.511617=T_c^{3D}$ (black circles). Dispersion of the magnetization
data is increased in the random dilution case and dispersion of the
concentration is increased in the thermal disposition. Inset shows the
histograms for the values of the normalized magnetization in both cases. The
upper case corresponds to random dilution and the lower to thermal dilution.
Note the difference of the shapes resulting in different values for the
normalized square widths.}
\label{fig2}
\end{figure}

The evolution with $L$ of the full scattering plot ($M$. vs. $c$) for the
critical thermal case is presented on Fig. 3. Note how the evolution with
increasing $L$ of the ''cloud'' of points reduces the dispersion in the
values of $c$, and the ''cloud'' tends to that for the canonical case, as
should be according to \cite{AharonyIII}. This behavior is also appreciable
for random dilution, even when the decrease in the dispersion on $c$ is not
so easily seen, due to the restriction to $p=0.5$ only (see Inset in Fig.3).
A more accuracy statistical study will be needed to investigate
quantitatively the behavior of $R_M$ with the dimension $L$ of our system, and,
in agreement with Ref. 13 higher values of $L$ will be needed to find total
universality of behavior between canonical and grand canonical constrains.

The main result of this study is the substantial improvement in the
self-averaging using a thermal distribution of vacancies for all $L$ values
considered under the ''grand- canonical'' constrain with $p=0.5$ and $c\geq
0.5$. The normalized squared width $R_M$ of the magnetization is always
around one order of magnitude bigger for the random case than for the
critically thermally diluted disposition of vacancies (see Fig.4).

\begin{figure}[h]
\epsfig{width=0.95 \linewidth,figure=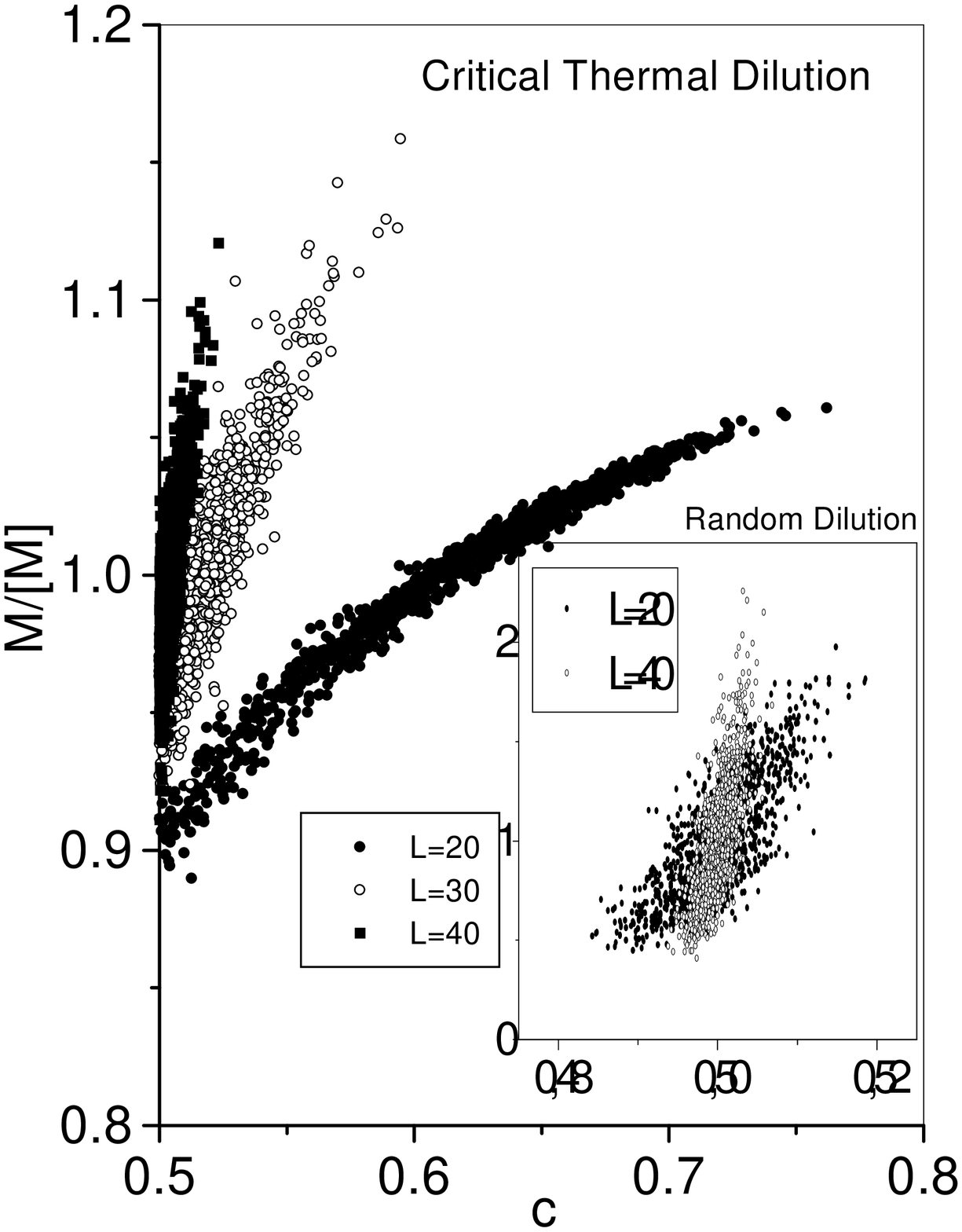} 
\vspace{0.0truecm}
\caption{Scattered plot of the concentration of spins ($c$) versus the
normalized magnetization per spin $M/[M]$ for critical thermal dilution at
different lengths ($L$), $L=20$ (black circles), $L=30$ (open circles) and $%
L=40$ (black squares). We find $[M]^{thermal}\approx 0.9$ for $L=20$, $%
[M]^{thermal}\approx 0.7$ for $L=30$ and $[M]^{thermal}\approx 0.6$ for $%
L=40 $. Note the decrease in the dispersion of $c$ as $L$ increases. This
behavior means that, as $L$ increases, the system tends to a constant value
of $c=0.5$ corresponding to the canonical constrain. Inset shows the same
scattered plot but now for the random case with $p=0.5$ and $L=20$ (black
circles) and $L=40$ (open circles).}
\label{fig3}
\end{figure}

Intuitively, an equilibrium thermal distribution of vacancies at criticality
implies a clustered distribution of both spins and vacancies for a given
distribution level ($p,c$). This implies an increase in the relative number
of bonds between neighboring spins, and therefore an increase of the
ordering and a subsequent decrease of the normalized square width for the
magnetization.

We can summarize our results as follows: using the Monte Carlo approach we
have investigated diluted ferromagnets using a new method to specify the
vacancy distribution which allows the ''thermal'' clustering of both the
magnetic atoms and magnetic vacancies under specific ''thermal'' conditions.
This kind of thermal dilution gives results totally equivalent to those
obtained with the usual random dilution method when the order-disorder
distribution of the vacancies is obtained away from criticality ({\bf %
hipercritical} conditions $\theta >>T_c^{3D}$) but it gives strongly
different results at {\bf critical} conditions ($\theta =T_c^{3D}$).

\begin{figure}[h]
\epsfig{width=0.95 \linewidth,figure=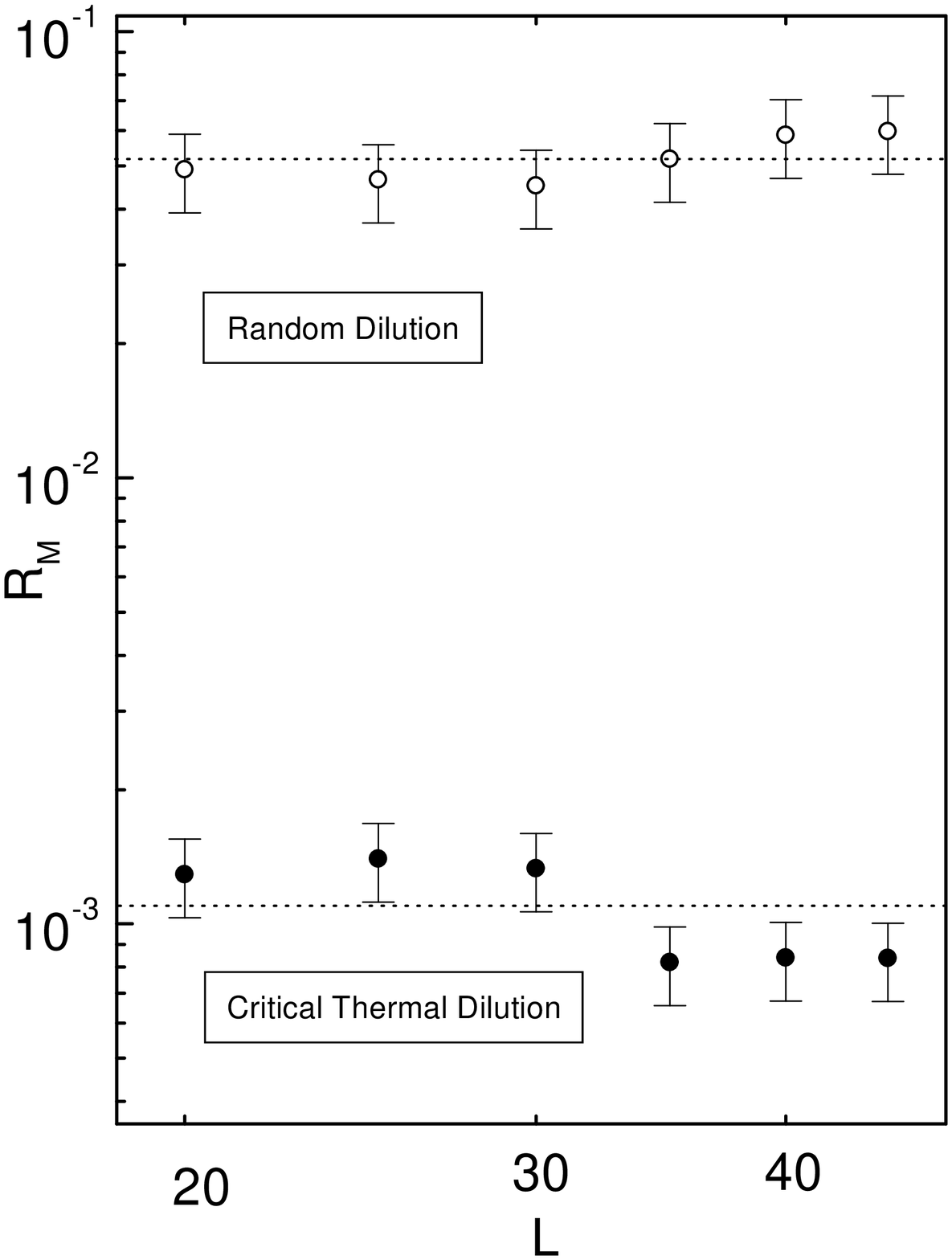} 
\vspace{0.0truecm}
\caption{Log-Log plot of the normalized squared width of the magnetization $%
R_M$ vs. the lenght $L$ of our system for the ''grand canonical'' case.
Results are shown for the random dilution with $p=0.5$ and $c\geq 0.5$ (open
circles) and for the critically thermally dilution (black circles).}
\label{fig4}
\end{figure}

In conclusion, our Monte Carlo results in critically thermally diluted (as
opposed to randomly diluted) ''grand-canonical'' samples with L increasing,
show substantial self-averaging improvement of the magnetization in three
dimensional highly diluted ($p=0.5$) Ising ferromagnets.

We gratefully acknowledge financial support from DGCyT through grant PB96-
0037.

\end{document}